\documentclass{article}
\usepackage{ijcai24}

\usepackage{times}
\usepackage{soul}
\usepackage{url}
\usepackage[hidelinks]{hyperref}
\usepackage[utf8]{inputenc}
\usepackage[small]{caption}
\usepackage{graphicx}
\usepackage{amsmath}
\usepackage{amsthm}
\usepackage{booktabs}
\usepackage{bigdelim, makecell} 
\usepackage{algorithm}
\usepackage{algorithmic}
\usepackage[switch]{lineno}
\usepackage{amssymb}
\usepackage{tikz}
\usepackage{cleveref}
\usepackage{multirow}
\usepackage{caption}
\usepackage{subcaption}
\usepackage{mathtools} 

\DeclareMathOperator*{\argmax}{argmax}

\DeclareSymbolFont{CMsymbols}{OMS}{cmsy}{m}{n} 
\DeclareSymbolFont{CMlargesymbols}{OMX}{cmex}{m}{n}
\DeclareMathDelimiter{\myrbrace}{\mathopen}{CMsymbols}{"67}{CMlargesymbols}{"09} 

\title{Learning Fair Cooperation in Mixed-Motive Games with Indirect Reciprocity}

\author{
Martin Smit \and Fernando P. Santos\\
\affiliations
Informatics Institute, University of Amsterdam\\
\emails
\{j.m.m.smit, f.p.santos\}@uva.nl
}

\begin{document}

\maketitle

\begin{abstract}
    Altruistic cooperation is costly yet socially desirable.
    As a result, agents struggle to learn cooperative policies through independent reinforcement learning (RL).
    Indirect reciprocity, where agents consider their interaction partner's reputation, has been shown to stabilise cooperation in homogeneous, idealised populations.
    However, more realistic settings are  comprised of heterogeneous agents with different characteristics and group-based social identities.
    We study cooperation when agents are stratified into two such groups, and allow reputation updates and actions to depend on group information.
    We consider two modelling approaches: evolutionary game theory, where we comprehensively search for social norms (i.e., rules to assign reputations) leading to cooperation and fairness; and RL, where we consider how the stochastic dynamics of policy learning affects the analytically identified equilibria.
    We observe that a defecting majority leads the minority group to defect, but not the inverse.
    Moreover, changing the norms that judge in- and out-group interactions can steer a system towards either fair or unfair cooperation.
    This is made clearer when moving beyond equilibrium analysis to independent RL agents, where convergence to fair cooperation occurs with a narrower set of norms.
    Our results highlight that, in heterogeneous populations with reputations, carefully defining interaction norms is fundamental to tackle both dilemmas of cooperation and of fairness.
\end{abstract}

\section{Introduction}

Cooperation is a fundamental research topic across disciplines \cite{rand_human_2013,fehr_nature_2003}.
While cooperative populations tend to thrive, individuals are tempted to act selfishly, receiving the benefits from the cooperation of others without exerting the effort themselves.
The conundrum underlying this interaction is evident if we formally translate it into the so-called \textit{donation game}, where a donor decides whether to pay a cost $c$ to offer a benefit $b$ to a recipient.
Assuming $b>c>0$, this simple interaction illustrates the ubiquitous social dilemma of altruistic cooperation.
These dilemmas are known as mixed-motive games as they combine the principles of competitive (i.e., zero-sum games) and cooperative interactions. 
Understanding how to engineer cooperation in mixed-motive settings is a fundamental scientific challenge \cite{pennisi_how_2005,rand_human_2013} and a key frontier in AI research \cite{paiva_engineering_2018,dafoe_cooperative_2021,conitzer_foundations_2023,fatima_learning_2024}.

In distributed multi-agent systems, research has focused on the design of autonomous systems where cooperation is stable \cite{genesereth_cooperation_1986}.
In such contexts, it is fundamental to understand how adaptive agents can learn to cooperate over time in a decentralised way \cite{claus_dynamics_1998}.
The cooperation mechanisms observed in human societies \cite{rand_human_2013} have inspired formal methods to stabilise cooperation in groups of artificial agents.

One particularly effective mechanism to sustain cooperation among humans is indirect reciprocity (IR) \cite{nowak_evolution_2005,santos_complexity_2021,okada_review_2020}.
Within such a framework, agents are assumed to discriminate and provide benefits based on the social standing of others; this mechanism relies on the availability of reputations.
The rules that determine how such reputations are updated (so-called \textit{social norms}) encapsulate the moral judgements of what constitutes a good or a bad action \cite{ohtsuki_how_2004}. 

Despite success in human populations, the application of IR in systems of learning agents is technically challenging.
As social norms assess the ``goodness'' of every action in every possible context, the number of possible norms grows combinatorially in the number of actions and states \cite{santos_social_2018}.
Moreover, small differences in otherwise similar norms can have unpredictable effects on reputation dynamics.
This makes predicting which norms will lead to a cooperative system a difficult task.
A central challenge is therefore determining how reputations should be assigned for cooperation to be maximised.
Previous work has shown that only a small set of social norms are able to stabilise cooperation in  populations of homogeneous agents \cite{ohtsuki_how_2004}.

Additionally, it is common for groups to exist or emerge in a population, possibly as a byproduct of existing reputation systems \cite{rosenblat_discriminating_2017,gross_rise_2019}, and for agents to consider group affiliation alongside reputations when making decisions.
Notably, when social norms governing reputations display in-group bias or out-group prejudice, even group-blind decisions through reputations can lead to inequality.
Interactions in online marketplaces, where reputations are key, offer a paradigmatic example.
The short-stay rental platform Airbnb, for instance, displays ratings and profile pictures for both hosts and guests.
Recent findings show that
guests with a distinctly African American name have a 12\% lower acceptance rate \cite{edelman_racial_2017}.

Although we observe cooperation dependent on the joint effect of reputation and group identity, the interaction of the two mechanisms has not yet been formalised.
In this paper, we fill this gap by answering whether independent agents that consider both reputations and group affiliations can learn to cooperate in an equitable manner despite biased social norms.
%
%
To answer this question, we employ two models which provide complementary perspectives: one focuses on analytically identifying norms that stabilise cooperative and fair states; the other focuses on how populations of reinforcement learning agents can reach the stable states identified.
As detailed below, we find that fairness and cooperation can be achieved if the right norm is chosen to judge actions.

\subsection{Structure of Paper and Contributions}
After introducing related literature on cooperation in mixed-motive games and indirect reciprocity (\textbf{\autoref{sec:related-work}}), subsequent sections offer the main contributions of our paper:
\begin{itemize}
    \item \textbf{Formalisation of a new model to systematically study cooperation and fairness under indirect reciprocity (\autoref{sec:model}).}
    We introduce group identities into an existing evolutionary game theory model to formalise differing treatment of in and out-group interactions by social norms.
    Our model allows us to study the evolutionary stability of cooperative and fair strategies under different norms \textbf{(\ref{sec:model:stab-anal})} and learning dynamics in a population of identical, but independent, Q-learning agents  \textbf{(\ref{sec:model:rl})}.
    \item \textbf{Stability analysis of norm-strategy combinations (\autoref{sec:egt-findings}).}
    We show that if the majority-identity defects, then the minority-identity cannot sustain cooperation even amongst themselves, but the inverse does not hold.
    Moreover, social norms that favour in-group interactions can sustain relatively high fairness and cooperation given a ``counterweight'' strategic bias.
    We go on to explore the effectiveness of well-studied norms when paired up to judge in-group/out-group interactions.
    \item \textbf{Analysis of cooperation and fairness dynamics of under Q-learning and group-structured populations (\autoref{sec:rl-findings}).}
    We investigate the impact of RL on learning cooperative and fair policies under specific norms.
    We show that, although cooperation decreases on average in an independent RL setting, agents  can learn fair cooperation.
    Despite this, we show that reaching fair/cooperative states when the benefits of cooperation are low requires an initial fraction of cooperative agents.
\end{itemize}

Finally, in \textbf{\autoref{sec:conclusion}}, we discuss the limitations of our results and possible directions for future work. \footnote{Appendix and code available at: \url{www.github.com/sias-uva/indirect-reciprocity}.
An extended abstract of this paper appears in the Proceedings of AAMAS'24 \cite{smit_fairness_2024}.} 

\section{Related Work}\label{sec:related-work}

\subsection{The Cooperation Dilemma}
Understanding human cooperation is a fundamental research topic across disciplines \cite{pennisi_how_2005,fehr_nature_2003,rand_human_2013}.
In AI, there is a growing interest in designing artificial agents to be cooperative and generate cooperation in others.
In a recent commentary \cite{dafoe_cooperative_2021}, the authors argue that AI requires ``social understanding'' and the ability to cooperate in order to achieve success in tasks that require complex interactions such as navigating pavements, financial markets, and online communication.
Many tasks that AI engage with also require cooperation with humans or other AI. Recent works have explored mechanisms to help enable cooperation.
The proposed methods include agents with inequality-aversion \cite{hughes_inequity_2018}, rewarding causal influence \cite{jaques_social_2019}, self-play \cite{anastassacos_cooperation_2021}, gifting \cite{lupu_gifting_2020} or introducing non-adaptive, pro-social agents \cite{santos_evolution_2019,anastassacos_cooperation_2021,guo_facilitating_2023}.



\subsection{Social Norms}
Previous work explores how coordination techniques and social governance can influence the autonomy of agents in a system and, for example, change the system's levels of cooperation.
Social norms implemented in computational systems can help in this endeavour \cite{savarimuthu_norm_2011}.
Two main classes of social norms are identified by \cite{villatoro_social_2010}: ``essential'' norms that seek to solve cooperation dilemmas and collective action problems \cite{griffith_why_2010,peleteiro_exploring_2014} and ``conventional'' – used to establish a convention, solving coordination dilemmas \cite{shoham_emergence_1997,sen_emergence_2007,morales_automated_2013}.
Our work focuses on \textit{essential} norms.
It is also common to divide norms into top-down ``legalist'' approaches, in which norms are designed offline and imposed by a central authority and bottom-up ``interactionist'' approaches, where norms are emergent phenomena (e.g., as defined in \cite{haynes_engineering_2017}).
Recent work explores the bottom-up creation of norms via so-called norm entrepreneurship \cite{anavankot_norm_2023} or observing public sanctions  \cite{vinitsky_learning_2023}. Our contribution involves both: while we apply norms top-down, their effectiveness is computed via a bottom-up process, where strategies evolve over time.

\subsection{Indirect Reciprocity in Multi-Agent Systems}
Indirect reciprocity (IR) has been proposed as a mechanism to elicit cooperation among reinforcement learning agents. Anastassacos \textit{et al.}~\shortcite{anastassacos_cooperation_2021} examine whether agents with private social norms learned through Q-learning can reach a socially optimal consensus on how reputations should be interpreted and updated.
To aid this learning, the researchers propose seeding the population with non-learning agents and through introspective self-play.
They find that a combination of both mechanisms can sustain cooperation. Differently from Anastassacos \textit{et al.}~\shortcite{anastassacos_cooperation_2021}, here agents play a \textit{one-sided} donation game as opposed to a \textit{two-sided} donation game (a prisoner's dilemma), and we shift learning from the norm space to the strategy space.
These decisions were made to align our paper with the extensive pre-existing IR literature.

While the asymmetrical nature of our game merely makes learning slightly less consistent due to delayed rewards, the shift from learning norms to strategies fundamentally changes the focus of the papers.
In their paper, the goal is to internalise the reputation mechanism and examine its effects on learning cooperation.
In contrast, we take the norm to be exogenous, introduce another variable upon which that agents can discriminate (group identity), and see how inequality can emerge in spite of cooperation.

\subsection{Group-structured Populations}
Some prior works studied IR in populations where agents explicitly belong to groups, through the lens of evolutionary game theory.
In this domain, Kessinger \textit{et al.}~\shortcite{kessinger_evolution_2023} assume that different groups might use different social norms and focus on the effect of different information broadcasting mechanisms, whereby information about individuals can spread only between members of the same groups or publicly (as in traditional models).
Contrary to the setting we explore here, Kessinger \textit{et al.}~\shortcite{kessinger_evolution_2023} assume that strategies only discriminate based on reputations and not group identity.
The authors find that in such systems, cooperation ultimately depends on the rate of in/out-group interactions, and cooperation can collapse if information remains within the same groups.

In a more recent work, Stewart and Raihani~\shortcite{stewart_group_2023} study how stereotypes might be formed through group reciprocity: the authors find that stereotyping can lead to \textit{negative judgement bias}, in which individuals become pessimistic about the willingness of out-group members to cooperate.
Although these works study dynamics of IR under reinforcement learning \cite{anastassacos_cooperation_2021} and dynamics of reciprocity associated with group identities \cite{stewart_group_2023}, the combination of indirect reciprocity, group identity and reinforcement learning remains under-explored.
In this paper, we propose a model that contributes to fill this gap.











\section{Model}\label{sec:model}
In this paper, a well-mixed population of agents, stratified into two groups, interact by playing a donation game.
In the donation game, a player is either a donor or a recipient.
The donor must decide whether to pay a cost to confer a benefit to their partner, and we assume that $b > c > 0$.

In the one-shot game, the dominant strategy for the donor is to not donate, i.e. to defect.
To encourage cooperation, we consider reputations and social norms.
Social norms encode the rules of society that confer agents with a corresponding reputation based on whether these rules are followed or broken.
In our model, these rules can depend on the action taken by the donor, the current reputation of the potential recipient (their ``goodness''), and whether the two agents are in the same group (their ``sameness'').
Following prior works on indirect reciprocity, all of these inputs are binary: a donor has two actions, reputations are either ``good'' or ``bad'', and agents can be in the same or a different group.

\begin{table}[thb]
    \centering
    \begin{tabular}{@{}llllll@{}}
     \addlinespace[-\aboverulesep]
     \cmidrule[\heavyrulewidth]{1-5}
    Donor action & 0 & 0 & 1 & 1 & \rdelim\}{2}{*}[ Input bits]\\ 
    Recipient reputation & 0 & 1 & 0 & 1 \\ \cmidrule{1-5}
    \textit{Shunning} (SH) & 0 & 0 & 0 & 1 & \rdelim\}{4}{*}[ Reputation]\\
    \textit{SternJudging} (SJ) & 1 & 0 & 0 & 1 \\
    \textit{ImageScoring} (IS) & 0 & 0 & 1 & 1 \\ 
    \textit{SimpleStanding} (SS) & 1 & 0 & 1 & 1 \\ \addlinespace[-\aboverulesep]
     \cmidrule[\heavyrulewidth]{1-5}
    \end{tabular}
    \caption{
        Social norms assign a reputation given different combinations of actions and recipient reputations.
        The input bits above represent the context in which an action takes place. By convention, 1 means Good/Cooperate and 0 means Bad/Defect.
        In our model, a norm is composed of two four-bit norms: one to judge in-group interactions and one to judge out-group interactions.
    }
    \label{tab:norms}
\end{table}

Although we study $2^8$ different social norms, in Table~\ref{tab:norms} we give examples of commonly studied norms.
One is \textit{SternJudging}, explored in detail by \citeauthor{pacheco_sternjudging_2006} \shortcite{pacheco_sternjudging_2006}, which deems that it is good to defect against a bad agent or cooperate with a good one, and that doing the opposite action in either case is bad.
Another is \textit{SimpleStanding}, which says that the only bad thing to do is to defect against a good agent \cite{panchanathan_tale_2003}.
Neither of these norms take into account the group relation of the agents involved, but a norm that judged in-group interactions with \textit{SternJudging} and out-group interactions with \textit{SimpleStanding} would imply a greater degree of strictness when judging interactions where both agents are members of the same group.
This specific norm, which could be called \textit{Stern/Stand}, judges in and out-group interactions differently, so we refer to it interchangeably as \textit{unfair} or \textit{discriminatory} and the others mentioned as \textit{fair} or \textit{group-agnostic}.
Fair and unfair norms are mutually exclusive.

We assume that agents can make execution errors.
An agent who intended to cooperate will sometimes defect with some probability $\epsilon$.
Following previous works, agents have zero probability of cooperating when intending to defect.
Similarly, we assume that third-party observers using social norms to assign reputations can also err and assign the opposite reputation than intended, with some probability $\delta$.

We assume a public reputation scheme where reputations are common knowledge among all agents.
Although we do not explicitly model the process of reputation diffusion, public reputations can exist due to gossiping about what happened over an efficient communication network or a central judge observing interactions and broadcasting reputations.

The way that agents decide how to act is determined by their strategy.
Strategies define, for each combination of reputation and in/out-group, a corresponding action.
As such, the space of strategies consists of functions $\sigma$ such that
\begin{equation}
    \sigma:\{0, 1\}^2 \to \{0, 1\}
\end{equation}
In the Q-learning model (see below), the Q-table holds Q-values associated with each action for each binary pair of information.
We refer to the (effective) strategy of an agent as the strategy that would result from applying the $\argmax$ function to the Q-table on each possible input pair.

Some notable strategies include \textit{AllD}, which unconditionally refuses to donate, \textit{Disc}, which conditionally cooperates with good agents and defects against bad ones, and \textit{AllC}, which unconditionally donates.
Unfair strategies may instead, for example, play \textit{AllC} with in-group members and play \textit{Disc} with everyone else.

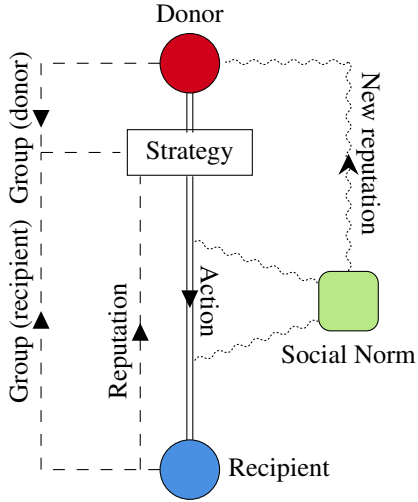
\begin{figure}[tb]
    \centering
    \begin{tikzpicture}[x=0.75pt,y=0.75pt,yscale=-1,xscale=1]
        
        \draw  [fill={rgb, 255:red, 208; green, 2; blue, 27 }  ,fill opacity=1 ] (115,40) .. controls (115,31.72) and (121.72,25) .. (130,25) .. controls (138.28,25) and (145,31.72) .. (145,40) .. controls (145,48.28) and (138.28,55) .. (130,55) .. controls (121.72,55) and (115,48.28) .. (115,40) -- cycle ;
        \draw  [fill={rgb, 255:red, 74; green, 144; blue, 226 }  ,fill opacity=1 ] (115,245) .. controls (115,236.72) and (121.72,230) .. (130,230) .. controls (138.28,230) and (145,236.72) .. (145,245) .. controls (145,253.28) and (138.28,260) .. (130,260) .. controls (121.72,260) and (115,253.28) .. (115,245) -- cycle ;
        \draw    (131.5,97) -- (131.5,230)(128.5,97) -- (128.5,230) ;
        \draw [shift={(130,163.5)}, rotate = 270] [fill={rgb, 255:red, 0; green, 0; blue, 0 }  ][line width=0.08]  [draw opacity=0] (8.93,-4.29) -- (0,0) -- (8.93,4.29) -- cycle    ;
        \draw  [dash pattern={on 0.75pt off 0.75pt}]  (209.8,39.4) .. controls (208.13,41.07) and (206.47,41.07) .. (204.8,39.4) .. controls (203.13,37.73) and (201.47,37.73) .. (199.8,39.4) .. controls (198.13,41.07) and (196.47,41.07) .. (194.8,39.4) .. controls (193.13,37.73) and (191.47,37.73) .. (189.8,39.4) .. controls (188.13,41.07) and (186.47,41.07) .. (184.8,39.4) .. controls (183.13,37.73) and (181.47,37.73) .. (179.8,39.4) .. controls (178.13,41.07) and (176.47,41.07) .. (174.8,39.4) .. controls (173.13,37.73) and (171.47,37.73) .. (169.8,39.4) .. controls (168.13,41.07) and (166.47,41.07) .. (164.8,39.4) .. controls (163.13,37.73) and (161.47,37.73) .. (159.8,39.4) .. controls (158.13,41.07) and (156.47,41.07) .. (154.8,39.4) .. controls (153.13,37.73) and (151.47,37.73) .. (149.8,39.4) -- (147.8,39.4) -- (147.8,39.4) ;
        \draw  [dash pattern={on 0.75pt off 0.75pt}]  (210,39.4) .. controls (211.67,41.07) and (211.67,42.73) .. (210,44.4) .. controls (208.33,46.07) and (208.33,47.73) .. (210,49.4) .. controls (211.67,51.07) and (211.67,52.73) .. (210,54.4) .. controls (208.33,56.07) and (208.33,57.73) .. (210,59.4) .. controls (211.67,61.07) and (211.67,62.73) .. (210,64.4) .. controls (208.33,66.07) and (208.33,67.73) .. (210,69.4) .. controls (211.67,71.07) and (211.67,72.73) .. (210,74.4) .. controls (208.33,76.07) and (208.33,77.73) .. (210,79.4) .. controls (211.67,81.07) and (211.67,82.73) .. (210,84.4) .. controls (208.33,86.07) and (208.33,87.73) .. (210,89.4) .. controls (211.67,91.07) and (211.67,92.73) .. (210,94.4) .. controls (208.33,96.07) and (208.33,97.73) .. (210,99.4) .. controls (211.67,101.07) and (211.67,102.73) .. (210,104.4) .. controls (208.33,106.07) and (208.33,107.73) .. (210,109.4) .. controls (211.67,111.07) and (211.67,112.73) .. (210,114.4) .. controls (208.33,116.07) and (208.33,117.73) .. (210,119.4) .. controls (211.67,121.07) and (211.67,122.73) .. (210,124.4) .. controls (208.33,126.07) and (208.33,127.73) .. (210,129.4) .. controls (211.67,131.07) and (211.67,132.73) .. (210,134.4) .. controls (208.33,136.07) and (208.33,137.73) .. (210,139.4) .. controls (211.67,141.07) and (211.67,142.73) .. (210,144.4) -- (210,144.9) -- (210,144.9) ;
        \draw [shift={(210,85.65)}, rotate = 90] [fill={rgb, 255:red, 0; green, 0; blue, 0 }  ][line width=0.08]  [draw opacity=0] (10.72,-5.15) -- (0,0) -- (10.72,5.15) -- (7.12,0) -- cycle    ;
        \draw  [fill={rgb, 255:red, 184; green, 233; blue, 134 }  ,fill opacity=1 ] (195,151) .. controls (195,147.69) and (197.69,145) .. (201,145) -- (219,145) .. controls (222.31,145) and (225,147.69) .. (225,151) -- (225,169) .. controls (225,172.31) and (222.31,175) .. (219,175) -- (201,175) .. controls (197.69,175) and (195,172.31) .. (195,169) -- cycle ;
        \draw  [dash pattern={on 4.5pt off 4.5pt}]  (55,85) -- (95,85) ;
        \draw  [dash pattern={on 4.5pt off 4.5pt}]  (55,245) -- (55,85) ;
        \draw [shift={(55,165)}, rotate = 90] [fill={rgb, 255:red, 0; green, 0; blue, 0 }  ][line width=0.08]  [draw opacity=0] (8.93,-4.29) -- (0,0) -- (8.93,4.29) -- cycle    ;
        \draw  [dash pattern={on 4.5pt off 4.5pt}]  (115,245) -- (55,245) ;
        \draw  [dash pattern={on 4.5pt off 4.5pt}]  (115,40) -- (55,40) ;
        \draw  [dash pattern={on 4.5pt off 4.5pt}]  (55,40) -- (55,85) ;
        \draw [shift={(55,72.5)}, rotate = 270] [fill={rgb, 255:red, 0; green, 0; blue, 0 }  ][line width=0.08]  [draw opacity=0] (8.93,-4.29) -- (0,0) -- (8.93,4.29) -- cycle    ;
        \draw  [dash pattern={on 4.5pt off 4.5pt}]  (105,245) -- (105,97.1) ;
        \draw [shift={(105,171.05)}, rotate = 90] [fill={rgb, 255:red, 0; green, 0; blue, 0 }  ][line width=0.08]  [draw opacity=0] (8.93,-4.29) -- (0,0) -- (8.93,4.29) -- cycle    ;
        \draw    (131.5,55) -- (131.5,75)(128.5,55) -- (128.5,75) ;
        \draw  [dash pattern={on 0.75pt off 0.75pt}]  (132.5,130) .. controls (134.61,128.95) and (136.19,129.48) .. (137.24,131.59) .. controls (138.29,133.7) and (139.87,134.24) .. (141.98,133.19) .. controls (144.09,132.14) and (145.67,132.67) .. (146.72,134.78) .. controls (147.77,136.89) and (149.35,137.42) .. (151.46,136.37) .. controls (153.57,135.32) and (155.15,135.85) .. (156.2,137.96) .. controls (157.25,140.07) and (158.83,140.61) .. (160.94,139.56) .. controls (163.05,138.51) and (164.63,139.04) .. (165.68,141.15) .. controls (166.73,143.26) and (168.31,143.79) .. (170.42,142.74) .. controls (172.53,141.69) and (174.11,142.22) .. (175.16,144.33) .. controls (176.21,146.44) and (177.79,146.98) .. (179.9,145.93) .. controls (182.01,144.88) and (183.59,145.41) .. (184.64,147.52) .. controls (185.69,149.63) and (187.27,150.16) .. (189.38,149.11) .. controls (191.49,148.06) and (193.06,148.59) .. (194.11,150.7) -- (195,151) -- (195,151) ;
        \draw  [dash pattern={on 0.75pt off 0.75pt}]  (132.5,190) .. controls (133.58,187.91) and (135.17,187.4) .. (137.26,188.48) .. controls (139.36,189.56) and (140.95,189.05) .. (142.02,186.95) .. controls (143.11,184.86) and (144.7,184.35) .. (146.79,185.43) .. controls (148.89,186.51) and (150.48,186) .. (151.55,183.9) .. controls (152.63,181.81) and (154.22,181.3) .. (156.31,182.38) .. controls (158.4,183.46) and (159.99,182.95) .. (161.07,180.86) .. controls (162.14,178.76) and (163.73,178.25) .. (165.83,179.33) .. controls (167.92,180.41) and (169.51,179.9) .. (170.6,177.81) .. controls (171.68,175.72) and (173.27,175.21) .. (175.36,176.29) .. controls (177.46,177.37) and (179.05,176.86) .. (180.12,174.76) .. controls (181.2,172.67) and (182.79,172.16) .. (184.88,173.24) .. controls (186.98,174.32) and (188.57,173.81) .. (189.65,171.71) .. controls (190.73,169.62) and (192.32,169.11) .. (194.41,170.19) -- (195,170) -- (195,170) ;
        
        \draw (130,15) node [anchor=center][inner sep=0.75pt]   [align=center] {Donor};
        \draw (175,245) node [anchor=center][inner sep=0.75pt]   [align=center] {Recipient};
        \draw (210,187) node [anchor=center][inner sep=0.75pt]   [align=center] {Social Norm};
        \draw (45,165) node [anchor=center][inner sep=0.75pt]  [rotate=-270] [align=center] {Group (recipient)};
        \draw (45,72.5) node [anchor=center][inner sep=0.75pt]  [rotate=-270] [align=center] {Group (donor)};
        \draw (140,160) node [anchor=center][inner sep=0.75pt]  [rotate=-90] [align=center] {Action};
        \draw (95,180) node [anchor=center][inner sep=0.75pt]  [rotate=-270] [align=center] {Reputation};
        \draw (220,85.65) node [anchor=center][inner sep=0.75pt]  [rotate=270] [align=center] {New reputation};
        
        \draw (98.5,73.5) rectangle (161.5,96.5) node[pos=.5] {Strategy};
    \end{tikzpicture}
    \caption{
    An illustration of an interaction between two agents which is observed by a third party using a social norm to update reputations.
    The donor's strategy determines the action taken based on the agents' relation to each other, and the reputation of the donor.
    The third party observes the action and context to assign the donor a new reputation based on a fixed social norm.
    }
    \label{fig:interaction-diagram}
\end{figure}

An example interaction is shown in Figure~\ref{fig:interaction-diagram}, in which the donor evaluates the recipient's reputation and group relation according to their strategy and takes the corresponding action.
A third party observes the action taken by the donor, and its context, and assigns the donor a new reputation.




\subsection{Evolutionary Stability}\label{sec:model:stab-anal}

Given the model introduced above, a natural question one can pose is: Given a norm, which strategies are more likely to be played by agents in the long-run?
One way this can be answered is with evolutionary game theoretical (EGT) tools.
A key concern of EGT is that of the evolutionary stability of strategies: as agents play the donation game, their payoffs rise and fall based on their strategy and its impact on their reputation; a (resident) strategy is evolutionarily stable if agents using an alternative (mutant) strategy are unable to achieve higher average payoffs and invade.

Assume that strategies $S_i$ have proliferated the entire incumbent population of each group $i$, and that the population is governed by social norm $N$.
By calculating the expected payoff of a player of each group playing their strategy $S_i$, we can determine whether a mutant strategy in any group could outperform the incumbents of that group.

To introduce IR, we assume a timescale separation between the evolution of reputations and the arrival of mutants, i.e. allowing reputations to converge before introducing mutants.
This allows us to derive and analytically solve the differential equations that give the long-term proportions of good agents in each group.
Our line of reasoning is identical to that of Ohtsuki and Iwasa~\shortcite{ohtsuki_how_2004}, with the added complication of groups with interdependent reputations.
The full derivation is available in Section A of the appendix.

After the reputations of players has stabilised, by introducing the utilities associated with each action, we can consider the long-run average payoffs of each population.
Given fixed reputations, the average player from any group has a fixed probability of cooperating and being cooperated with every time they partake in an interaction according to each population's strategy, the relative size of each population, and the reputations of each population.

If we associate to each group $i$ a benefit received \(b_i\) and cost of cooperating \(c_i\), then we can determine the long term payoffs \(U_i\) of each group as
\begin{equation}
    U_i = \sum_{j=1}^2 \mathbb{P}(i \xleftrightarrow{\text{int.}} j)\left(b_i\mathbb{P}(j \xrightarrow{\text{don.}} i) 
 - c_i\mathbb{P}(i \xrightarrow{\text{don.}} j) \right)
\end{equation}
where \(\xleftrightarrow{\text{int.}}\) means ``interacts with'', \(\xrightarrow{\text{don.}}\) means ``donates to'', and the probabilities of donating are derived from the stationary reputations, strategies, and error rates of each group.


By doing a similar (yet simpler) derivation of mutant reputations and payoffs, one can determine if the combination of norms and strategies \((N, S_1, \dots, S_K)\) is an \emph{evolutionarily stable state} (ESS).

Stronger than the traditional Nash equilibrium, a strategy $S_I$ is evolutionarily stable on the condition that if any alternative strategy $S_M$ arises in a group that \(S_I\) has proliferated and that the proportion of agents playing this alternative is sufficiently small, then this alternative strategy will perform worse than the incumbent strategy $S_I$ and die out.
We say that a combination of norm and strategies is an ESS if all of its strategies are evolutionarily stable.
If all strategies in a combination are AllD, then the combination is trivially an ESS because any alternative strategy that cooperates with anyone would immediately be worse off.

While EGT and stability analysis informs which strategies are stable under each norm, we need to understand 1) how prevalent each equilibrium point is and 2) whether learning agents converge to a certain ESS.
For both purposes, we use multi-agent reinforcement learning.





\subsection{Reinforcement Learning}\label{sec:model:rl}
We model agents as independent (tabular) Q-learners.
Due to the asymmetry in the donation game, while agents always learn from every interaction, the donor can only be negatively reinforced by possibly donating, and recipients can only be positively reinforced by possibly receiving a donation.
Nevertheless, no matter the interaction, the Q-values of both parties decay due to the learning rate.
We assume that recipients attributes whatever donations they may receive (or lack thereof) to their most recently taken action as a donor.


Formally, if agent $i$ meets agent $j$ who has $x$ relation to agent $i$ and $y$ reputation, then the action taken by $i$ ($a^\ast$) is determined by the equation

\begin{equation}\label{eq:decision}
    a^{\ast} = \argmax_{a \in \{0, 1\}}Q_i[x, y, a]
\end{equation}
where $Q_i \in \mathbb{R}^3$ is the Q-table of agent $i$ and actions $0$ and $1$ correspond to defection and cooperation respectively.

In doing so, agent $j$ will receive payoff $a^{\ast}b$ and agent $i$ will pay cost $a^{\ast}c$.
Agent $j$ will attribute this payoff to the last action they took ($\hat{a}$) in context $\hat{x}$ and $\hat{y}$

\begin{gather}\label{eq:learning}
    Q_i^{\text{t+1}}[x, y, a^\ast] \leftarrow (1-\alpha) Q_i^t[x, y, a^\ast] - \alpha a^{\ast} c\\
    Q_j^{\text{t+1}}[\hat{x}, \hat{y}, \hat{a}] \leftarrow (1-\alpha) Q_j^t[\hat{x}, \hat{y}, \hat{a}] + \alpha a^{\ast}b
\end{gather}




In our model, strategies take into account the in/out-group relation, and not which specific group another agent is from.
When combined with imbalances in group size, benefit-to-cost ratio, and error rates, this minimal two-group model can capture the most pertinent fairness results while minimising complexity and the computational work necessary.

\subsection{Metrics of Social Desirability}
We measure the performance of a system with two metrics:
\begin{itemize}
    \item The \textit{cooperativeness} is the probability that, in a uniformly sampled interaction, the donor will cooperate.
    \item The \textit{fairness} is the ratio between the average payoffs of the worst off and best off group.
    This is akin to the \textit{demographic parity} ratio from supervised learning. 
\end{itemize}

\subsection{Experimental Setup}\label{sec:model:setup}
In all the following experiments, analytical or agent-based, the reader may assume the following parameters were used unless otherwise stated:
The rate of agent execution errors and judgement execution errors is relatively rare at \(1\%\), and the benefit-to-cost ratio in our analytical model is \(5\) with \(c=1\), \(b=5\).
This represents a scenario where cooperation is highly beneficial (high \(b/c\) ratio), yet defection is still a dominant strategy (\(c>0\)). 
Furthermore, the majority group comprises 90\% of the population, and agents in different groups are functionally identical.
These choices allow us to isolate the effects of discrimination and the effects of inherent differences between agents' in different groups.
We refer readers to the appendix for a discussion of alternative parameter setups (group size, benefits, errors).

We run our RL experiments with a population of 50 agents (45 in the majority group and 5 in the minority group).
We fix the exploration rate \(\mu\) and learning rate \(\alpha\) to 0.1.
Each simulation runs for \(250,000\) interactions and we run each simulation 50 times with a different seed.
The source code for this paper (models, experiments, and figures) is available on \href{https://github.com/sias-uva/indirect-reciprocity}{GitHub}.

\section{Results}
\subsection{Stability Analysis}\label{sec:egt-findings}
First, we evaluate the stability of all possible combinations of a norm and a strategy in each group, of which there are \(2^{16}\) possibilities.
We distinguish between defective strategies (Always defect/\textit{AllD}), strategies ignoring group identity (Group-agnostic), and strategies discriminating based on group-identity (Discriminatory).
By considering these categories we aim at developing an intuition for the possibility that fair and unfair cooperation emerge which, respectively, relies on stable Group-agnostic and Discriminatory strategies.

\begin{figure}[tb]
    \centering\includegraphics[scale=0.55]{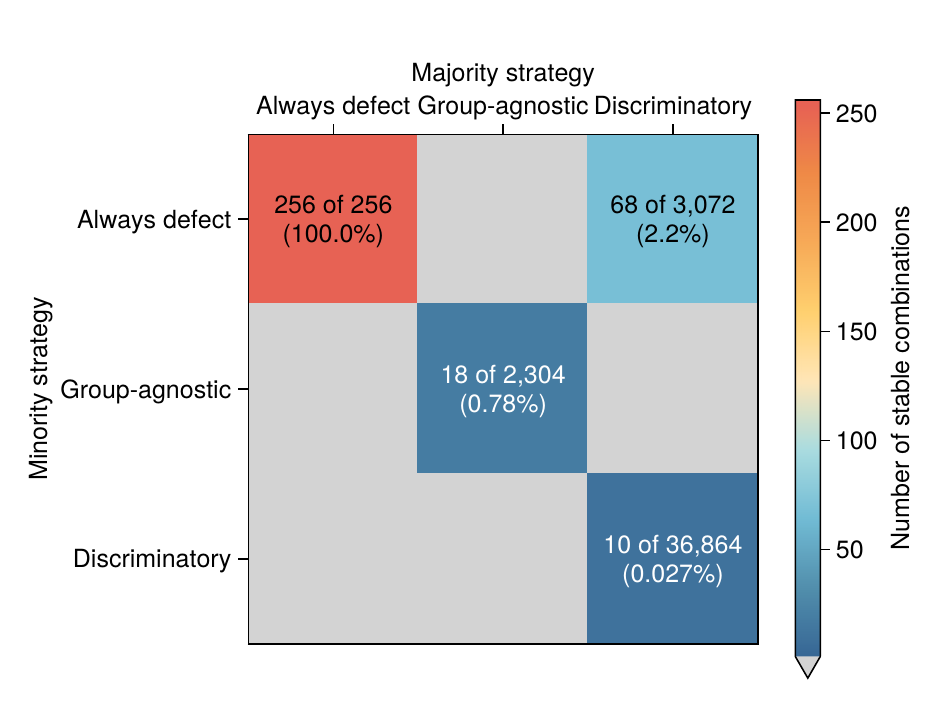}
    \caption{
        We evaluate all \(2^{16}\) NSS (Norm-Strategy-Strategy) combinations of norms and strategies used by the two groups.
        We categorise every stable NSS according to the cooperative and discriminatory nature of strategies involved: strategies can defect regardless of the opponent type (Always defect), ignore group identity (Group-agnostic) or discriminate based on group identity (Discriminatory).
        We observe that group-agnostic strategies are unable to coexist with any other type of strategy, and a defecting majority playing \textit{AllD} leads a minority to defect -- but not the inverse.
        Parameters used: \(b/c=5\), error rate \(= 0.01\).
        In the appendix (Figure~1) we confirm that the number of stable states remains unchanged for a wide range of error rates and \(b/c\).
    }
    \label{fig:stable-table}
\end{figure}

Figure~\ref{fig:stable-table} shows which strategies are stable under this setup.
We observe that group-agnostic strategies by one group cannot sustainably coexist with discriminatory strategies in another,
which means that a whole group ignoring group identities can make such identity discrimination unstable for everyone.
Furthermore, we observe that the majority group can dictate whether cooperation can be stable:  
if the majority group unconditionally defects, there is no stable strategies where the minority cooperates; 
the inverse does not hold.
As can be also seen in Figure~\ref{fig:stable-table}, conditional cooperation from the majority group is a prerequisite for cooperation in the minority group.

%
\begin{figure}[tb]
    \centering
    \includegraphics[scale=0.55]{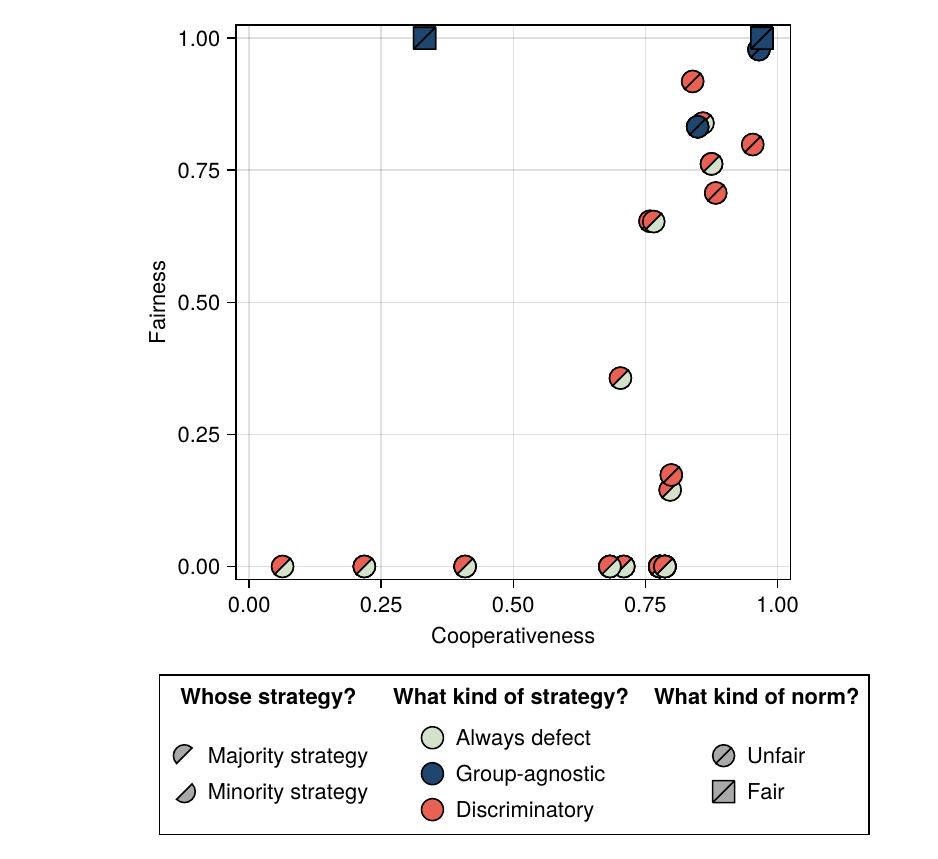}
    \caption{
        Fair norms (plotted as squares), which assign reputations independently of group identities, lead to fairness, but not necessarily high levels of cooperation (top-left).
        High cooperation and fairness can be stable with both fair and unfair norms (top-right quadrant).
        Parameters used: same as Figure~\ref{fig:stable-table}.
    }
    \label{fig:coop-fair-scatter}
\end{figure}
In Figure~\ref{fig:coop-fair-scatter} we explicitly compute the level of cooperativeness and fairness for each stable combination of strategies, given any norm fixed in the population.
We observe that norms assigning reputations independently of group identities, lead to high levels of fairness, but not necessarily high levels of cooperation.
Unfair norms, on the other hand, can lead to a cooperative yet unfair system, where a minority group always defects and the majority only cooperates with in-group members.
Surprisingly, however, unfair norms can also sustain highly cooperative and fair systems, through the stability of group-agnostic strategies.
This means that the observed levels of cooperation and fairness in a system can not trivially be inferred from the fairness level of a norm.
Due to execution errors, the cooperativeness of a system can never achieve a perfect score of 100\% even when all players always \textit{intend} to cooperate.
However, several norms are able to achieve equivalent levels of cooperation and fairness, and these are superimposed in the very top-right of the figure.
When we inspect the common bits between these norms, we find that, regardless of whether the norm treats in-group and out-group the same, the two must be treated by either \textit{SimpleStanding} or \textit{SternJudging} (see Table~\ref{tab:norms}). 



\subsubsection{Evaluating Well-known Social Norms}\label{sec:famous}

\begin{figure}[tb]
    \centering
    \includegraphics[scale=0.55]{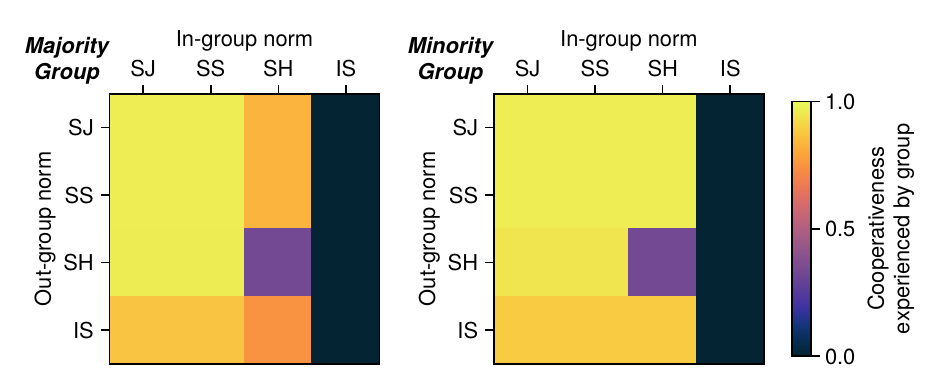}
    \caption{
    We use abbreviations from Table~\ref{tab:norms} to refer to the norms.
    Diagonal entries are previously studied ``fair'' norms.
    Overall, the majority group is most affected by the in-group norm, and vice-versa for the minority group.
    In-group-\textit{ImageScoring} causes a total cooperation breakdown, whereas in-group-\textit{Shunning} is most impacted when paired with out-group-\textit{Shunning}, as agents have fewer ways to recover reputation.
    }
    \label{fig:famous-norms-heatmap}
\end{figure}

A number of ``leading'' norms, which can consistently stabilise cooperation, have been identified in previous works \cite{ohtsuki_how_2004,ohtsuki_leading_2006}.
These norms agree that cooperating with good individuals is good, and defecting against good individuals is bad.
In Figure~\ref{fig:famous-norms-heatmap}, we see the analytical levels of cooperation produced by each combination of norms when used to judge in- and out-group interactions.
We find that, as in previous works \cite{ohtsuki_global_2007}, \textit{SternJudging} and \textit{SimpleStanding} (defined in  Table~\ref{tab:norms}) stabilise cooperation.
However, with the addition of groups, we are also able to examine cooperation when interactions with outsiders are judged according to a different social norm.
We observe that, as long as in-group interactions are judged according to \textit{SternJudging} or \textit{SimpleStanding}, out-group interactions can be evaluated according to other norms (e.g., the strict norm \textit{Shunning}) without significantly decreasing cooperation.
In fact, we observe that one can even use \textit{ImageScoring} to judge out-group interactions and recover high levels of cooperation.
This is remarkable given that this norm is simple and only relies on information about an action --- a property that can be instrumental in settings where the reputations of out-group members are not widely accessible.

\subsection{Learning Fair Cooperation}\label{sec:rl-findings}
\begin{figure}[thb]
    \centering
    \includegraphics[scale=0.55]{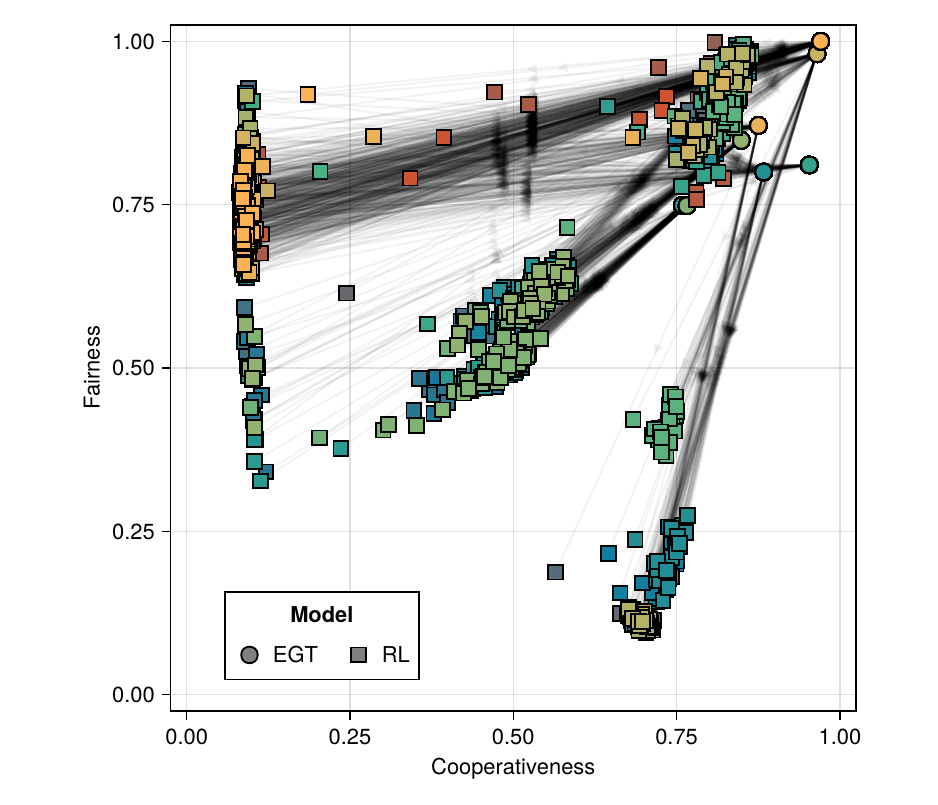}
    \caption{
    Some norms can sustain fair and cooperative equilibria, yet may be ineffective at guiding a population of independent RL agents to converge towards such states even with an elevated \(b/c\) ratio of 10, a trend which this figure demonstrates.
    We use different colours to represent different highly cooperative and fair norms (from the top-right quadrant of Figure~\ref{fig:coop-fair-scatter}).
    Circles represent EGT results, while squares represent RL results.
    For some norms, RL agents are unable to converge to the strategies that theoretically form an equilibrium, which leads to lower levels of fairness, cooperation, or both.
    Some norms, such as \textit{SternJudging} (and variations) are impacted very little, and thus indicate that strictness is required for independent RL agents to sustain fairness and cooperation.
    In Section C of the appendix, we explore which strategies are learned to lead to these outcomes.
    }
    \label{fig:egt-to-rl-scatter}
\end{figure}

\begin{figure}[thb]
    \centering
    \includegraphics[scale=0.55]{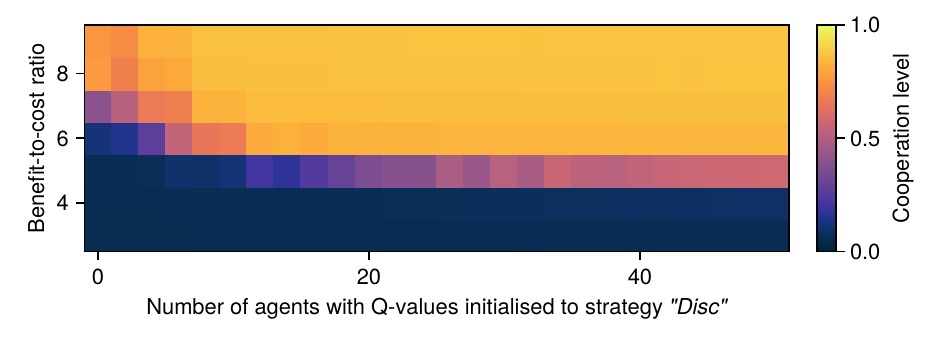}
    \caption{
    \textit{SternJudging} can sustain high levels of cooperation in a multi-agent RL setting, but for stricter dilemmas (lower \(b/c\) ratios), convergence to a cooperative equilibrium is less likely and highly dependent on the initial distribution of Q-values.
    To demonstrate this, the x-axis shows that some agents were initialised with Q-values corresponding to the socially optimal strategy: ignoring groups and cooperating only with good individuals.
    }
    \label{fig:cooperation-heatmap}
\end{figure}

The previous (EGT) stability analysis informs which norms stabilise cooperation and, as a result, be effective in sustaining fair cooperation in groups of heterogeneous agents.
Identifying norms that theoretically stabilise fairness and cooperation is a computationally attractive way of filtering norms that, when implemented in a system of learning agents, can possibly sustain fair cooperation.
Learning fair cooperation in a finite population of adaptive agents requires, however, that agents are able to converge to a desirable equilibrium (which is naturally not trivial).
Here, we explicitly model a population of independent RL agents.
Figure~\ref{fig:egt-to-rl-scatter} shows that even with a higher benefit-cost ratio ($b/c=10$), some norms that were previously predicted to stabilise fair cooperation are not able to do so consistently under RL.
Importantly, Figure~\ref{fig:egt-to-rl-scatter} reveals that moving from an EGT to an RL analysis can affect one or both of cooperation and fairness, and to a different extent, depending on the norm in question.
Furthermore, this effect is worsened when the dilemma is more difficult, and cooperation depends on the initial Q-values, as shown in Figure~\ref{fig:cooperation-heatmap}.

%
%
Although some norms might lead to fair and cooperative stable states, they differ in the size of the attraction basins that lead to such states.
Furthermore, finite populations of learning agents are subject to stochastic effects that might prevent reaching fair and cooperative states. 


Despite the challenges of converging to cooperative/fair equilibria, indirect reciprocity constitutes a promising mechanism to sustain cooperation and fairness in independent RL domains.
We conclude that social norms such as \textit{SternJudging} (and close variants) can be applied to multi-agent reinforcement learning domains to, in group-structured populations, to ensure both cooperation and fairness in the long-run.
The effectiveness of this norm can be augmented by resorting to seeding agents, as previous works suggest \cite{anastassacos_cooperation_2021}).
We observe that a combination of indirect reciprocity and seeded agents can be harnessed to, not only sustain cooperation \cite{anastassacos_cooperation_2021}, but also supporting fairness in a heterogeneous group-structured population.

\section{Conclusion}\label{sec:conclusion}

In this paper, we have shown that indirect reciprocity allows for fair cooperation among group-structured agents.
For this to happen, one has to judiciously select social norms; norms play a large part in determining the stability and learnability of policies leading to a fair and cooperative outcome.
We showed that a large variety of norms and strategies can be stable, with varying levels of cooperation and fairness.
We showed that well-known norms (like \textit{SternJudging}) perform well when agents adapt through reinforcement learning, being able to consistently achieve near to the idealised levels of cooperation and fairness predicted by our analytical model.

By using the minimal and generic donation game, we offer a proof of concept whose results may inform the application of indirect reciprocity to more elaborate multi-agent systems and motivates future work along several dimensions:
While we make the common EGT assumption of a well-mixed population, a preference for interacting with similar individuals has been observed among humans \cite{fu_evolution_2012}.
Furthermore, groups in our model are static.
It has been shown that ad hoc group formation may occur in spatial mixed-motive models \cite{bara_role_2023} or in complex networks \cite{gross_rise_2019} where diverse local conventions can evolve \cite{hu_local_2017}; investigating learning dynamics with a changing interaction structure or group labels could inform how to sustain cooperation and fairness in scenarios where group membership is dynamic. 

Prior work focuses on norm emergence and dynamics \cite{de_understanding_2017,savarimuthu_norm_2011}.
In the context of indirect reciprocity, work in this domain indicates that decentralised reputation systems with private information  can hinder cooperation \cite{hilbe_indirect_2018}, especially when the norm itself must also be learned \cite{anastassacos_cooperation_2021}. This requires extra mechanisms to retain cooperation under private reputations \cite{krellner_pleasing_2022,kawakatsu_mechanistic_2024}.
It may also be the case in a fully decentralised setting that the norm changes over time, something that may lead to an effective norm being chosen \cite{pacheco_sternjudging_2006} or a complete breakdown in cooperation \cite{xu_cooperation_2019}.
It would be interesting to explore how norms could be modified over time to maintain stable, fair cooperation between groups.

Despite these directions for future work, the model we formalise, and following comprehensive study, already sheds light on the advantages and challenges of indirect reciprocity in a minimal setting of a group structured population.
With this modelling approach, we offer a link between algorithmic fairness
-- typically considered in the context of supervised \cite{mehrabi_survey_2021} or unsupervised \cite{chierichetti_fair_2017} learning --
and  multi-agent systems where reputations exist and social dilemmas of cooperation need to be solved.
Our model provides a base framework to identify norms that, in such a context, sustain high levels of cooperation and fairness, thereby guaranteeing that universal cooperation is attained and parochialism avoided.

\section*{Acknowledgements}
Co-funded by the European Union (ERC, RE-LINK, 101116987).
Views and opinions expressed are however those of the author(s) only and do not necessarily reflect those of the European Union or the European Research Council. Neither the European Union nor the granting authority can be held responsible for them.

\bibliographystyle{named}
\bibliography{ijcai}

\end{document}


\maketitle

\counterwithin{figure}{section}
\appendix
\section{Derivation of Evolutionary Stability}
In this section, we derive the expected value for the proportion of good individuals in each population given a specific setup.
As agents' strategies (introduced formally below) take their interaction partner's reputation into account, the average ``goodness'' of an individual in either group is required to calculate the likelihood that any agent cooperates or defects in an arbitrary interaction.

With this expected cooperation level of each agent, we can derive whether any agent could unilaterally deviate from their strategy and improve their utility through some combination of cooperating less often and receiving more donations through improving their reputation.
As we make the standard evolutionary game theory (EGT) assumption that both groups are infinitely large, any unilateral deviation won't affect the other agents' reputations or utilities.
This mathematical convenience allows us to rapidly test every possible alternative strategy for agents in either group.

If no agents can improve their expected utility through unilateral deviation, then we say that the norm and combination of strategies found in each group is an evolutionarily stable state (ESS).

\subsection{Derivation of Reputation Dynamics}

Our derivation follows the setup and notation of Section~3.1, in the main text where agents are stratified into two groups representing proportions $p$ and $1-p$ respectively of the total well-mixed population, interacting through the donation game.
We assume that in each group, one strategy has entirely proliferated the population i.e. every agent in a group plays the same strategy.

A strategy is a well-defined function from the space of reputations (Good/Bad) and group relations (In/Out-group) to actions (Cooperate/Defect).
Given that these inputs and outputs are all binary, the space of strategies \(\Sigma\) consists of functions $\sigma$ with
\begin{equation}
    \sigma:\{0, 1\}^2 \to \{0, 1\}
\end{equation}
Similarly, a ``social norm'' $N$ (hereby just a ``norm'') is a function from the space of reputations, group relations, and actions to reputations:
\begin{equation}
    N:\{0, 1\}^3 \to \{0, 1\}
\end{equation}
In our paper, we assume that, when used as inputs, 0 means ``out-group'', ``bad reputation'', and ``defect'', while 1 means ``in-group'', ``good reputation'', and ``cooperate''.

In other works a norm may have different inputs.
For example in \cite{ohtsuki_leading_2006}, norms additionally take the donor's own reputation as an input.
Given that norms in our paper can treat in-group and out-group interactions entirely differently, each norm can be uniquely described by how it treats such interactions, and so in Table 1 of our main text, we give examples of well-known norms that could be ``combined'' to judge in-group and out-group interactions.

We additionally assume that agents make execution errors at a rate \(\epsilon\).
These execution errors randomly prevent agents from cooperating when they intend to.
In expectation, we can represent these errors as functions \(E_\epsilon\) such that if \(x\) is the probability of an agent to cooperate:
\begin{equation}
    E_\epsilon(x) = (1-\epsilon)x.
\end{equation}

Similarly, we assume that norms make assignment errors at a rate \(\delta\).
These assignment errors assign the inverse reputation than what was intended, and so if \(y\) is the intended reputation assignment, the actual outcome in expectation will be
\begin{equation}
    E_\delta(y) = (1-\epsilon)y + \epsilon(1-y).
\end{equation}

Let \(G_t = [G_M, G_m]_t \in [0, 1]^2\) be a vector of the proportion of good individuals in the majority and minority populations at (continuous) time $t$.
We aim to derive, and solve, an equation for \(\tfrac{G_t}{dt}\).

With interactions occurring uniformly with some arbitrary rate such that, over some time period \([t, t + \Delta)\), a proportion \(\Delta\) of all agents interact once as a donor and \(1 - \Delta\) do not interact as a donor at all.
Of this proportion \(\Delta\), fraction \(p\) of them will be majority-group agents and \(1-p\) will be minority-group agents.

Let's take the specific example of two majority players interacting, where the recipient has a good reputation.
Hence, the input to the strategy \(\sigma\) would be \([1, 1]\).
Due to the well-mixed assumption, the chance of this interaction occurring is \(p^2 \times G_M\).
The donor, comprised of strategy \(\sigma\) and execution error \(E_\epsilon\) with rate $\epsilon$ would take action
\begin{equation}
    x = (E_\epsilon \circ \sigma)([1, 1])
\end{equation}
(where \(\circ\) denotes function composition) which would then be judged with norm \(N\) having assignment error \(E_\delta\) at rate \(\delta\) leading to the expected new reputation of the donor to be:
\begin{align}
    y &= (E_\delta \circ N)([1, 1, x])\\
     &= (E_\delta \circ N)([1, 1, (E_\epsilon \circ \sigma)([1, 1])])
\end{align}

Let \(\sigma_M\) and \(\sigma_m\) be the strategies that have proliferated the majority and minority populations respectively.
Given that the strategy \(\sigma\) and norm \(N\) share two inputs, namely the reputation of the recipient and the group relation of the individuals, it is notationally convenient to define the functions \(J_M\) and \(J_m\) where
\begin{align}
    J_M(a, b) &= (E_\delta \circ N)([a, b, (E_{\epsilon_M} \circ \sigma_M)([a, b])])\\
    J_m(a, b) &= (E_\delta \circ N)([a, b, (E_{\epsilon_m} \circ \sigma_m)([a, b])]).
\end{align}
Note that the execution error rates can be different between groups; we do not assume that \(\epsilon_M = \epsilon_m\).
in our previous example of two majority agents meeting and the recipient being good, we can simply calculate that the expected reputation of the donor is \(J_M(1, 1)\).

The above interaction is just one of many that may occur.
To calculate the expected change in reputation of an individual from (for example) the majority group \(\Delta G_M\), we sum the expected reputation outcome of a specific interaction multiplied by the probability that this interaction will occur.
Specifically, we sum over meeting a member of the in or out group, and whether this agent is good or bad.

By enumerating all possible interactions as such, we can derive the expected change \(\Delta G_M\) in reputation in time \(\Delta t\) as follows.
For players who do not interact as a donor, their reputation is unchanged, and so vector \(G_t\) of them are good.
For those who do interact as a donor, they can interact with a majority or minority player with proportion \(p\) or \(1-p\), and this player will be good with proportion \(G_M\) or \(G_m\) respectively.
Following the same logic for the minority group, we can calculate the expected reputation of a majority or minority donor.
Therefore we can write:
\begin{align}\label{eq:GM(t+dt)}
    \begin{split}
        G_M(t+\Delta t) &= (1 - p\,\Delta t)G_M(t)\\
        &+ p\,\Delta t\left(\mathbb{E}\left[\text{Reputation after interacting}\right]\right)
    \end{split}
\end{align}

Then, by rearranging (\ref{eq:GM(t+dt)}) to get
\begin{equation}
    G_M(t + \Delta t) - G_M(t),
\end{equation}
on the LHS, dividing by \(\Delta t\) and letting \(\Delta t \to 0\),  then doing the same for \(G_m\), we can derive the corresponding system of differential equations for \(G_t\).
The form of these equations allows the system to be written as the matrix equation:

\begin{equation}\label{eq:dGdt}
    \frac{d}{dt}
    \begin{bmatrix}
        G_M \\
        G_m
    \end{bmatrix} =
    \begin{bmatrix}
        p & 0\\
        0 & (1-p)
    \end{bmatrix}
    \left(
    (A - I)
    \begin{bmatrix}
        G_M \\
        G_m
    \end{bmatrix}
    + 
    b\right)    
\end{equation}
where $I$ is the \(2 \times 2\) identity matrix and
\begin{equation}
    \begin{split}
    &A = \\
    &\begin{bmatrix}
        p(J_M(1, 1) - J_M(1, 0)) & (1-p)(J_M(0, 1) - J_M(0, 0))\\
        p(J_m(0, 1) - J_m(0, 0)) & (1-p)(J_m(1, 1) - J_m(1, 0))
    \end{bmatrix}, 
    \end{split}
\end{equation},
\begin{equation}
    \begin{split}
        &b =\begin{bmatrix}
            p\,J_M(0, 0) + (1-p) J_M(1,0)\\
            p\,J_m(0, 0) + (1-p) J_m(1,0)
        \end{bmatrix}.
    \end{split}
\end{equation}

\subsection{Proof of Existence and Uniqueness}

Given that (\ref{eq:dGdt}) is a \(2 \times 2\) matrix differential equation, it has a fixed point \(G^\ast\) given by
\begin{equation}\label{eq:Gstar}
    G^\ast=-(A-I)^{-1} b
\end{equation}
as long as \(A-I\) is invertible.
Moreover, this fixed point is stable if its eigenvalues have negative real part, which is equivalent in the \(2\times2\) case to the conditions that \(A-I\) has negative trace and positive determinant (see, for example, \cite{sanchez_ordinary_1968}).

Firstly, we see that each value on the diagonal of \(A\) is less than 1 and so the values on the diagonal of \(A - I\) are both negative, hence the trace is negative.

Secondly, taking a real \(2\times2\) matrix \(C\) of the form
\begin{equation}
    C = \begin{bmatrix}
        pa & (1-p)b\\
        pc & (1-p)d\\
    \end{bmatrix}
\end{equation}
where \(a,b,c,d \in [0,1]\), then
\begin{align}
    \det(C - I) &= (pa - 1)((1-p)d - 1) - p(1-p)bc\\
                      &= (1 - pa)(1 - (1-p)d) - p(1-p)bc
\end{align}
which is \textit{uniquely} minimised to \(0\) when \(a = b = c = d = 1\):
\begin{align}
    \det(C - I) &= (1 - p)(1 - (1-p)) - p(1-p)\\
                &= (1-p)p - p(1-p)\\
                &= 0
\end{align}
In this case, the original equation (\ref{eq:Gstar}) using \(A\) instead of \(C\) does not have a solution.
However, if we assume that agents always have a non-zero probability to make execution errors, then the functions \(J_M\) and \(J_m\) cannot output the values of 0 and 1.
Hence, the values of \(a, b, c\) and \(d\), which are comprised of the outputs of these functions, are strictly between 0 and 1.
So, with this assumption, can guarantee that the differential equation (\ref{eq:dGdt}) has a unique fixed point given by (\ref{eq:Gstar}) which is stable.
This implies that the reputations of players in any system will reach a unique equilibrium invariant of the initial conditions, for a fixed combination of strategies in the population.

\section{Strategy Dynamics and Evolutionary Stability}
\subsection{Payoffs of incumbents}
We assume timescale separation between the reputations and strategic updates.
This assumption means that reputations are fixed by the time strategies are updated, which allows us to calculate, given a strategy having proliferated each group, the likelihood that any agent will cooperate, their utility, and ultimately whether any agent can improve their utility by changing their strategy.

In the main text we introduce benefits \(b_i\) and costs \(c_i\) to group \(i \in \{1, 2\}\) in order to determine the expected utility \(U_i\) of each group as
\begin{equation}\label{eq:Ui}
    U_i = \sum_{j=1}^2 \mathbb{P}(i \xleftrightarrow{\text{int.}} j)\left(b_i\mathbb{P}(j \xrightarrow{\text{don.}} i) 
 - c_i\mathbb{P}(i \xrightarrow{\text{don.}} j) \right)
\end{equation}
where \(\xleftrightarrow{\text{int.}}\) means ``interacts with'' and \(\xrightarrow{\text{don.}}\) means ``donates to''.

While \(\mathbb{P}(i \xleftrightarrow{\text{int.}} j)\) can be calculated simply by multiplying together the proportions of the population that groups \(i\) and \(j\) represent, the values of \(\mathbb{P}(j \xrightarrow{\text{don.}} i)\) and \(\mathbb{P}(i \xrightarrow{\text{don.}} j)\) are dependent on the (stationary) reputations of the groups.

Take \(\mathbb{P}(j \xrightarrow{\text{don.}} i)\) with \(i=1\) and \(j=2\) i.e. we want to calculate the probability that a minority agent will donate to a majority agent.
Using our previous notation for the agent's strategy \(\sigma_m\), execution error \(E_{\epsilon_m}\), and the reputations \(G^\ast\), we can write

\begin{align}
    \mathbb{P}(j \xrightarrow{\text{don.}} i) &= \mathbb{P}(i \text{ is good}) \mathbb{P}(j \xrightarrow{\text{don.}} i \mid i \text{ is good})\nonumber\\
    &+ \mathbb{P}(i \text{ is bad}) \mathbb{P}(j \xrightarrow{\text{don.}} i \mid i \text{ is bad})\\
    &= G^\ast_M (E_{\epsilon_m} \circ \sigma_m)([0, 1]) \nonumber\\
    &+ (1 - G^\ast_M)\left[(E_{\epsilon_m} \circ \sigma_m)([0, 0])\right].
\end{align}

By performing such a calculation for every sum, we can use (\ref{eq:Ui}) to derive \(U = [U_1, U_2] = [U_M, U_m]\), the average utility/payoff of an agent in each group.

\subsection{Payoffs of mutant strategies}

Assume now that some agent in one of the groups changes their strategy from \(\sigma\) to \(\sigma^\prime\).
As we assume that populations are infinitely large, their small change has no bearing on the utilities of others, only themselves.
By changing their strategy, they will now have a different average reputation to the other members of their group.
This reputation can be derived similarly to that of the incumbents, but it is easier because \(G^\ast\), the reputations of the incumbents, is fixed due to timescale separation between reputational and strategic updates.

Let the current reputation of a mutant in group \(i\) of size \(p_i\) be \(\tilde{G}_i\), such that \(\tilde{G} = [\tilde{G}_M, \tilde{G}_m]\).
As before in \ref{eq:GM(t+dt)}, we can derive the equation
\begin{align}
\tilde{G}_i(t + \Delta t) &= (1 - p_i\,\Delta t)\tilde{G}_i(t)\\
    &+ p_i\,\Delta t\left(\mathbb{E}\left[\text{Reputation after interacting}\right]\right)
\end{align}
Where, as before, we split the expectation up into each group and good/bad individuals, then calculating the expected reputation of interacting with each of these combinations.

Subtracting \(\tilde{G}_i(t)\), dividing by \(\Delta t\) and taking the limit of $\Delta t \to 0$, we have the following differential equation with a similar, but simpler form to before:

\begin{align}\label{eq:dGtildedt}
    \frac{d}{dt}\tilde{G}_t =
    \frac{d}{dt}
    \begin{bmatrix}
        \tilde{G}_M \\
        \tilde{G}_m
    \end{bmatrix} =
    \begin{bmatrix}
        \tilde{G}_M \\
        \tilde{G}_m
    \end{bmatrix}
    +
    \left(
    \tilde{A}
    \begin{bmatrix}
        G^\ast_M \\
        G^\ast_m
    \end{bmatrix}
    + 
    \tilde{b}
    \right)
\end{align}

where, \(A\) and \(b\) are exactly as define before however the functions \(J_M\) and \(J_m\) are replaced with \(\tilde{J}_M\) and \(\tilde{J}_m\) respectively where the only difference between the two is that the incumbent strategy \(\sigma\) is replaced with the currently considered mutant strategy \(\sigma^\prime\).
Note that \(\tilde{A}\) is multiplied by \(G^\ast\) and not \(\tilde{G}\).

Setting \(\tfrac{d}{dt}\tilde{G}_t = 0\), the solution to (\ref{eq:dGtildedt}) can simply be read off as
\begin{equation}
    \tilde{G}^\ast=-\left(
    \tilde{A}
    \begin{bmatrix}
        G^\ast_M \\
        G^\ast_m
    \end{bmatrix}
    + 
    \tilde{b}
    \right).
\end{equation}

Applying (\ref{eq:Ui}) using the mutant reputations leads to the payoffs \(\tilde{U}\) for mutants in either group playing strategy \(\sigma^\prime\) instead of \(\sigma_M\) or \(\sigma_m\).

\subsection{Evolutionary Stability}
If a combination of norm $N$ and strategies \(\sigma_M\) and \(\sigma_m\) cannot be ``invaded'' by any mutant strategy \(\sigma^\prime \in \Sigma\setminus\{\sigma\}\) i.e. no player in any group can unilaterally deviate and strictly improve their utility, then the triple \((N, \sigma_M, \sigma_m)\) is an evolutionarily stable strategy (ESS).
Technically this is a weaker condition than the traditional definition of an ESS (for example, the one found in \cite{smith_logic_1973}), which stipulates that equality \textit{is} allowed unless mutants also do strictly better amongst themselves than incumbents do against mutants.
We do not consider this case due to floating point inaccuracies causing equality to be unreliable.

\section{Reinforcement Learning}
In this section, we clarify a number of details regarding the RL model and Q-learning algorithm introduced in Section~3.2 of the main paper.
We also present some additional results 

Firstly, while in many previous non-analytical approaches to indirect reciprocity such as \cite{pacheco_sternjudging_2006} a large number of interactions is done before any strategic updates take place, in our Q-learning approach our version of  strategic updates, updating Q-values, happens after every single interaction for the agents involved.

Secondly, reputation updates also take place after each interaction, as in the EGT model.
The fact that reputation updates and strategic updates happen at a similar timescale could be a reason for some of the discrepancies in results between the RL and EGT models.

Thirdly, as in the EGT model, agents are sampled completely randomly to interact with each other in a well-mixed way with no thought for who has or hasn't already interacted.
This means that, even though we use a population of 50, in 50 interactions it is exceedingly unlikely that all agents engage exactly twice in these interactions, once as a donor and once as a recipient.

\begin{figure}[tb]
    \centering\includegraphics[scale=0.55]{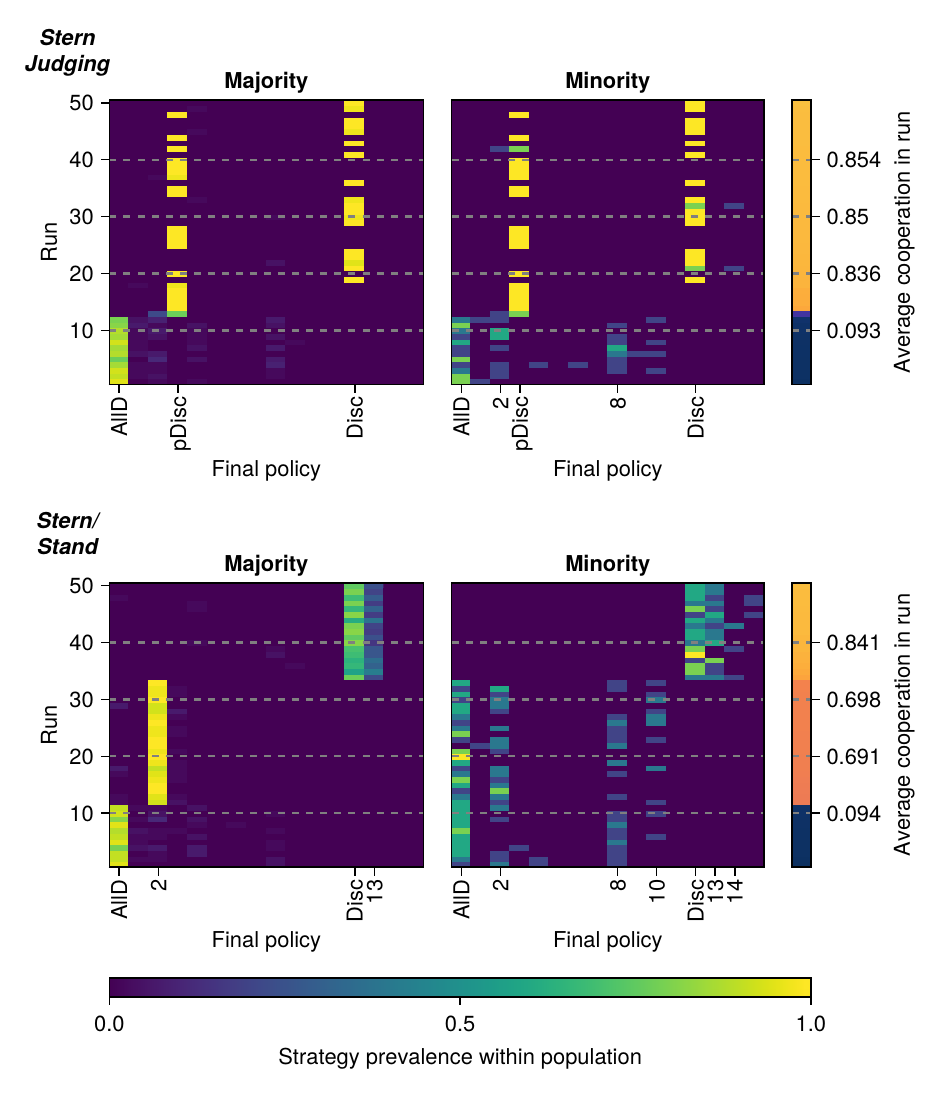}
    \caption{
        A horizontal slice of this figure indicates the prevalence of each strategy in each group, and the resulting cooperativeness, in a particular RL simulation, for a particular norm.
        We say an RL agent ``plays'' a certain strategy by taking the highest Q-value for each action.
        The slices are presented in increasing levels of cooperation, which is found as a colourbar to the right of each pair of heatmaps.
        While \textit{SternJudging} semi-consistently leads to high cooperation, reflected in main text Figure~6, \textit{Stern/Stand} (in-group \textit{SternJudging}, out-group \textit{SimpleStanding}) has three ``regimes'' that appear to occur roughly equally often.
        The x-axis only contains labels for strategies passing a threshold of frequency (\(25\%\)).
        Numbered/unnamed strategies are 2 (cooperate with bad-ingroup), 8 (cooperate with good-ingroup), 10 (cooperate with all ingroup), 13 (don't cooperate only with bad-ingroup), and 14 (don't cooperate only with bad-outgroup).
    }
    \label{fig:prevalence_heatmap}
\end{figure}

In Figure~\ref{fig:prevalence_heatmap} we examine the policies learned by RL agents in 50 different simulations.
As indicated in Figure~5 of the main text, \textit{SternJudging} is able to relatively consistently guide agents towards \textit{Disc} or \textit{pDisc}, which both lead to high levels of cooperation.

\section{Dilemmas of Varying Difficulty}
In the main text, we present Figure~2 concerning the number of stable combinations and Figure~3 concerning each stable combination's levels of cooperation and fairness.
These figures were generated using the parameters described in Section~3.4.

\begin{figure}[tbh]
    \centering\includegraphics[scale=0.55]{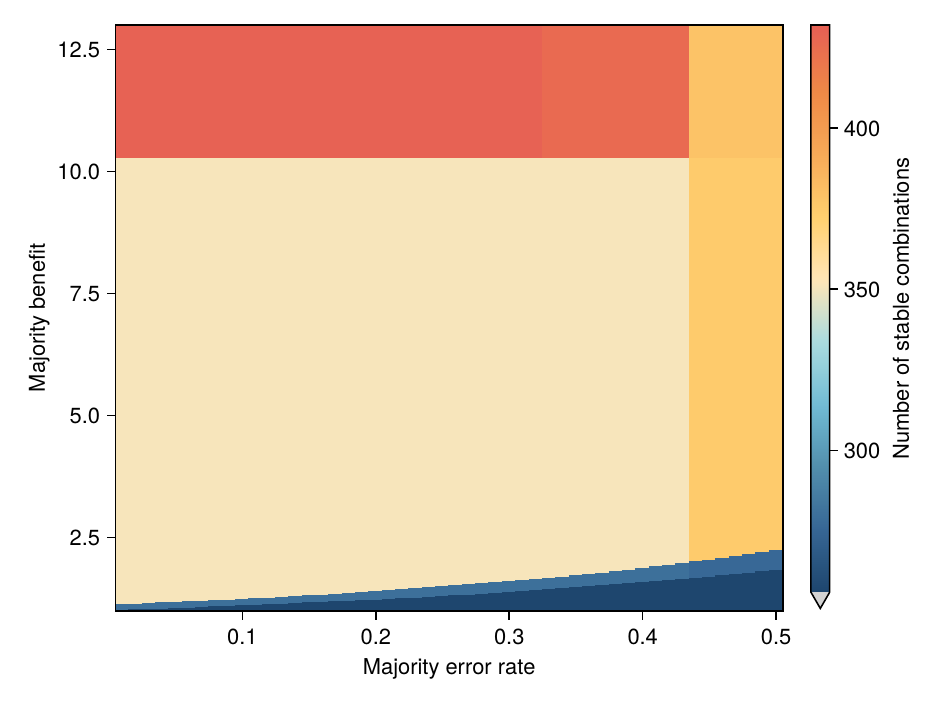}
    \caption{
        As the majority group's benefit from receiving a donation and error rate is changed, the number of stable combinations exhibits phase transitions where a number of combinations become stable or lose stability.
        At the bottom of the plot, we see cooperation dilemmas which are so difficult that the only stable combinations are those where no one cooperates.
    }
    \label{fig:phase-diagram}
\end{figure}

Varying these parameters would lead to more or less difficult cooperation dilemmas, and when doing so we found that the number of stable combinations can change.
Figure~\ref{fig:phase-diagram} shows that as difficulty increases (i.e., the cooperation benefit decreases and error rates increase in the majority group), the number of stable combinations changes.
Additionally, these changes exhibit ``phase transitions'' i.e. changing parameters has no effect until a certain threshold is reached, after which the stability of a large set of strategies is affected.
Our standard parameter setup from Section~3.4 is found in the largest area, which itself is surprisingly resilient, ranging all the way from a benefit-to-cost ratio of 1.2 to 10, and an error rate from (almost) 0 to \(0.43\).

\begin{figure}[tbh]
    \centering\includegraphics[scale=0.55]{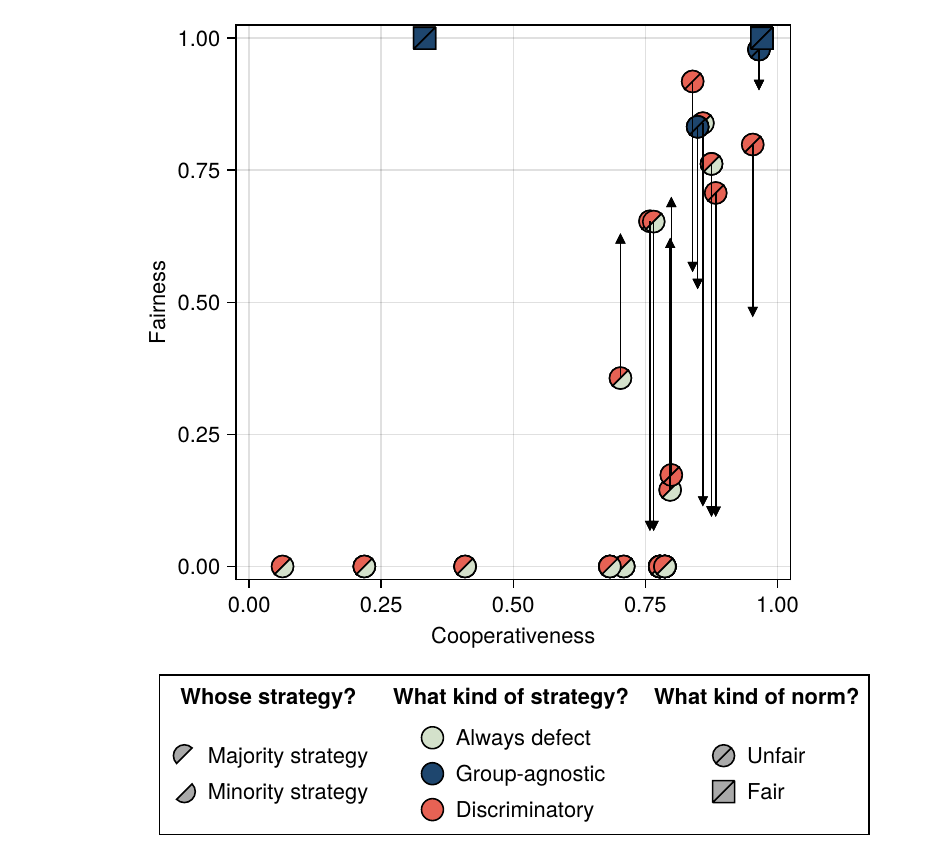}
    \caption{
        The markers are plotted where $b/c = 5$, and the arrows point to where $b/c = 1.25$.
        This is still inside the same region of the phase diagram, so all of the points are still stable, but the level of fairness is drastically different for a number of norms which feature non-zero between-group cooperation yet lack perfect fairness.
    }
    \label{fig:scatter-hard-dilemma-diff}
\end{figure}

In the analytical results of the main manuscript, we used a benefit/cost ratio of 5. If we change this to 1.25, almost within the ``no nontrivial stable combinations'' region of Figure~\ref{fig:phase-diagram}, we see in Figure~\ref{fig:scatter-hard-dilemma-diff} that the levels of cooperation are unchanged, and the levels of fairness go either up or down depending on the norm.
Some norms are unaffected at all, these are those that have either 0 or 1 fairness.

These findings are a consequence of our choice of metric to measure cooperation and fairness.
By definition, when lowering the benefit/cost ratio within the boundaries of the phase diagram, for no norms-strategy-strategy combinations will another strategy invade.
This implies that the current strategy is still optimal, and so agents will act in the same way in the same circumstances.
However, these circumstances (i.e. the prevalence of each group in the population, and the prevalence of good individuals in each group) also remain unchanged precisely because the strategies remain unchanged.
Given a fixed norm, reputations are determined by actions, which are determined by strategies.
Hence, reputations, and therefore the probability of each individual to cooperate, remains unchanged when the benefit/cost ratio changes within the boundaries of the phase diagram.

As cooperativeness is defined in our paper as ``the proportion of interactions that result in cooperation'', this remains unchanged when the benefit/cost ratio is changed.
Fairness, on the other hand, is defined as the ratio of payoffs between the best-off and worst-off group.
Given that the rate of cooperation in each group is unchanged, if one group had, for example, a high chance to receive cooperation and a low chance to cooperate, their payoff would be relatively much less impacted than a group that had the same chance to receive cooperation but a higher chance to cooperate.
Hence, under our definition of fairness, fairness can change even when rates of cooperation stay the same.

\section{Effect of Group Sizes}
We chose to focus on the case where the majority group comprised 90\% of the population.
With this decision, the majority agents would interact amongst themselves 90\% of the time, and minority amongst themselves only 10\% of the time.

In IR, agents have two, sometimes competing goals: sustaining reputation and maximising utility.
For successful norms like \textit{SternJudging}, the threat of unjustified defection is strong as agents can maintain their reputation by defecting against you.
The less successful norm \textit{Shunning} suffers from impossibly high standards: any interaction with a bad individual leads to a bad reputation.

In our two group setting, norms that distinguish in- and out-group interactions offer agents the opportunity to build up their reputation under a different norm, or to take advantage of a more lenient norm to raise their utility.
Consider norms where one part, either in- or out-group, is \textit{Shunning}.
Figure~4 in the main text shows us that, when paired with another norm, the cooperation level increases hugely compared to when it is paired with itself.
Furthermore, this change is not because its paired norm is more successful than \textit{Shunning} at sustaining cooperation as this applies even to \textit{In-Shunning/Out-ImageScoring}.

\begin{figure}[tbh]
    \centering\includegraphics[scale=0.55]{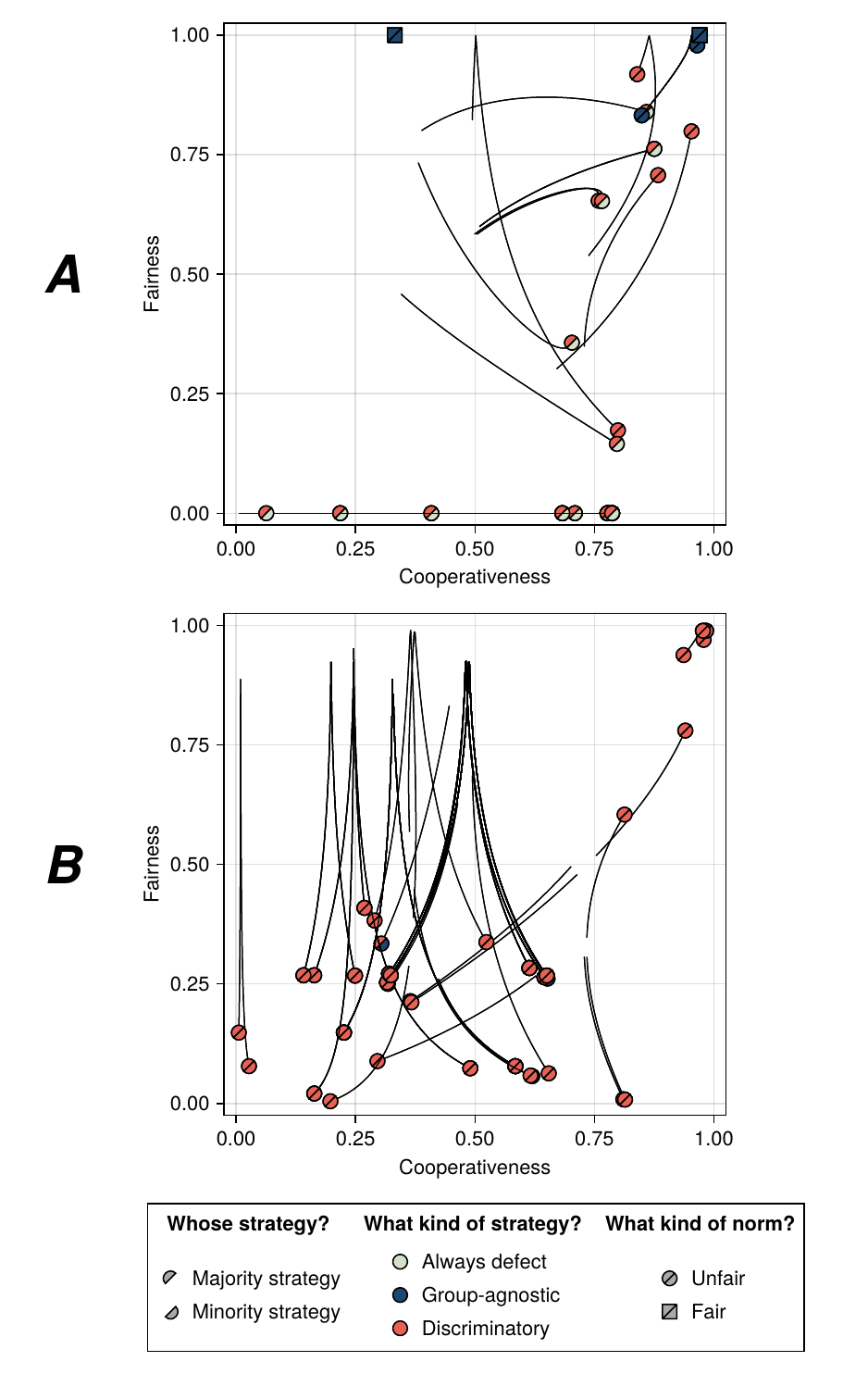}
    \caption{
        By examining the cooperation and fairness of different NSS combinations as the group size is varied, we notice a highly heterogeneous set of ``trajectories''.
        The group size varies from a slight majority 52\%, to a large majority 90\%.
        The largest majority for which the NSS combination is stable is marked with a shape, either a circle or square representing an unfair or fair norm respectively.
        The upper plot labelled \textbf{A} contains all NSS combinations that are stable at 90\% majority size, as found in Figure 3 of the main text.
        On the contrary, \textbf{B} contains all combinations that were not stable when the majority comprised 90\%.
    }
    \label{fig:groupsize-diagram}
\end{figure}

While the balance between reputation and utility is more complex in a setting with group relation dependent norms, in Figure~\ref{fig:groupsize-diagram} we vary the group size of the majority between 52\%, a very slight majority, and the value of 90\% used in our paper.
Notably, no fair norms that are stable at 52\% are unstable at 90\%, and we observe the same for NSS combinations where the minority group plays \textit{AllD}.
This highlights the difficulty of the scenario studied in our article and the importance of considering group size as a key part of modelling in future work.

\begin{table*}[thb]
\centering
\begin{tabular}{@{}llllllllll@{}}
\toprule
IN & ON & MIS & MOS & mIS & mOS & Cooperation (EGT) & Cooperation (RL) & Fairness (EGT) & Fairness (RL) \\ \midrule
Sh & Sh & Disc & Disc & Disc & Disc & 0.332 & 0.118  & 1.0   & 0.917 \\
Sh & SJ & Disc & Disc & Disc & Disc & 0.849 & 0.532  & 0.848 & 0.609 \\
Sh & IS & AllC & Disc & AllD & AllD & 0.766 & 0.456  & 0.748 & 0.502 \\
Sh & SS & Disc & Disc & Disc & Disc & 0.849 & 0.48   & 0.848 & 0.596 \\
SJ & Sh & Disc & Disc & Disc & Disc & 0.965 & 0.718  & 0.981 & 0.727 \\
SJ & SJ & Disc & Disc & Disc & Disc & 0.971 & 0.652  & 1.0   & 0.909 \\
SJ & IS & AllC & Disc & AllD & AllD & 0.875 & 0.599  & 0.871 & 0.541 \\
SJ & SS & Disc & Disc & Disc & Disc & 0.971 & 0.61   & 1.0   & 0.541 \\
IS & Sh & AllD & AllD & AllD & AllD & 0.0   & 0.106  & 1.0   & 0.727 \\
IS & SJ & AllD & AllD & AllD & AllD & 0.0   & 0.0948 & 1.0   & 0.74  \\
IS & IS & AllD & AllD & AllD & AllD & 0.0   & 0.0996 & 1.0   & 0.712 \\
IS & SS & AllD & AllD & AllD & AllD & 0.0   & 0.0929 & 1.0   & 0.737 \\
SS & Sh & Disc & Disc & Disc & Disc & 0.965 & 0.161  & 0.981 & 0.756 \\
SS & SJ & Disc & Disc & Disc & Disc & 0.971 & 0.132  & 1.0   & 0.747 \\
SS & IS & AllC & Disc & AllD & AllD & 0.875 & 0.108  & 0.871 & 0.741 \\
SS & SS & Disc & Disc & Disc & Disc & 0.971 & 0.0904 & 1.0   & 0.741 \\ \bottomrule
\end{tabular}
\caption{
We make the following abbreviations: N = norm, S = strategy, I/O = in-group/out-group, M/m = Majority/minority, EGT = The results came from the analytical evolutionary game theory model, RL = The results came from running the RL model 50 times and taking the average.
The strategy columns are the optimal set of strategies (in terms of highest cooperation) from the EGT model given an in-group and out-group norm.
Parameters used: $b/c = 10$, error rate = $0.01$
}
\label{tab:norm-performance}
\end{table*}

\section{Availability of code}
The code to produce all figures, tables, and values found in the paper and appendix is available under an MIT license on GitHub\footnote{Link to repository: \href{https://github.com/sias-uva/indirect-reciprocity/}{github.com/sias-uva/indirect-reciprocity}}.
The code is written in the Julia programming language \cite{bezanson_julia_2012} with figures in Makie \cite{danisch_makie_2021}.

\bibliographystyle{named}
\bibliography{ijcai24_appendix}